# An econophysical approach of polynomial distribution applied to income and expenditure

Elvis Oltean

e-mail: elvis.oltean@alumni.lboro.ac.uk

**Abstract:** Polynomial distribution can be applied to dynamic systems in certain situations. Macroeconomic systems characterized by economic variables such as income and wealth can be modelled similarly using polynomials. We extend our previous work to data regarding income from a more diversified pool of countries, which contains developed countries with high income, developed countries with middle income, developing and underdeveloped countries. Also, for the first time we look at the applicability of polynomial distribution to expenditure (consumption). Using cumulative distribution function, we found that polynomials are applicable with a high degree of success to the distribution of income to all countries considered without significant differences. Moreover, expenditure data can be fitted very well by this polynomial distribution. We considered a distribution to be robust if the values for coefficient of determination are higher than 90%. Using this criterion, we decided the degree for the polynomials used in our analysis by trying to minimize the number of coefficients, respectively first or second degree. Lastly, we look at possible correlation between the values from coefficient of determination and Gini coefficient for disposable income.

## 1. Introduction

Polynomials are known to explain the activity of dynamic systems in Physics, at least in certain situations. Considering that macroeconomic systems can be assimilated to dynamic systems, we apply polynomial distribution to income and expenditure on a pool of six countries. The horizon of applicability for polynomial distribution is vast, as so far were identified only disposable income and wealth which seem to obey this kind of probabilistic distribution [1]. In this paper we seek to explore further possibilities of applicability to other data from several other countries regarding mean income calculated as disposable income and expenditure. The countries analyzed so far are mostly very developed countries, implicating that they are characterized by high income. In order to test the applicability of the polynomial function, we succeeded to find data from countries which belong to very different categories. The diversity of these countries is about their different level of economic development, different macroeconomic results, different fundamental characteristics such as resources, model of development, level of income and so on.

## 2. Short Theoretical Framework and Literature Review

Modern approach of income and wealth distribution was done extensively by Yakovenko [2], [3], [4], [5] and Kusmartsev [6], [7], which was mostly about Maxwell-Boltzmann, Bose-Einstein, and lognormal (Gibrat) distributions. The first paper analyzing the applicability of Fermi-Dirac distribution to income and wealth [8] states that money distribution in an economy behaves similarly with the distribution of electrons in quantic systems.

Polynomial distribution came to the attention as dynamic systems (including the analogue macroeconomic ones) can be modelled using polynomials in certain cases when the dynamic systems have certain characteristics. Considering that short term evolutions are the analogues of a snapshot in a dynamic environment, polynomial distribution proved to be a robust distribution for income and wealth in macroeconomic systems. A property of this distribution consists of the applicability on the entire range of income and/or wealth including for upper income segment of population which is thought to be described solely by Pareto distribution.

## 3. Methodology

In this paper, we use our previous approach [1] to data regarding income and expenditure from other countries in order to test the degree of applicability on countries having very different characteristics from one another and also compared to the characteristics of the previously studied countries. Moreover, this is the first attempt of a paper to tackle the expenditure/consumption in a probabilistic manner by using distributions specific to Statistical Mechanics. Thus, we previously approached countries which fall in the category of developed countries (France, Finland, and Italy), characterized by high income and relevant exposure to crisis effects. The pool countries approached in this paper have a large diversity, thus enabling us to draw further conclusions regarding the applicability to other macroeconomic systems having very different characteristics and which were affected to a different extent by the most recent world economic crisis. This opens the way to a more comprehensive analysis of macroeconomic systems by using Statistical Physics distributions.

The countries we consider in our analysis have several different characteristics. Thus, from the point of view of the development the UK, Sweden, Brazil, and Singapore are developed countries, Philippine is a developing country, and Uganda is an underdeveloped one. From the point of view of income, the UK and Sweden are countries with high income, Singapore and Brazil are countries with middle income, while Philippine and Uganda are countries with low income.

From the point of view of economic growth and, indirectly, the extent to which the crisis affected the overall macroeconomic evolution, there are several categories. Thus, Philippine, Singapore, and Uganda were not affected by the economic crisis, Brazil's economic growth diminished but did not affect it to the point of falling in recession, while the UK and Sweden were affected to some extent. It is noteworthy that Brazil had different currencies during analyzed years (new cruzeiro and reais).

The data we use in this paper are expressed in deciles, which are fragmented by dividing the population ranged in increasing order of the values for their income. Deciles are divided equal parts of population that contain 10% of the population. Thus, the lowest decile of income is the first one and includes the poorest part of populations sorted according to their income. The highest decile of income (the tenth) includes the richest part of population and is believed to be described by Pareto distribution. The notion we employ to measure the income is the mean income. Mean income is the sum of all the individual income belonging to a certain decile divided to the number of people belonging to that decile.

The method we use to calculate probability of the population which has of income above a certain threshold is the cumulative probability distribution function (cdf). In order to have a better understanding, we present below the formula for continuous approximation

$$P(x) = \int_{x}^{\infty} p(t)dt \quad (1)$$

P represents fraction of population with income or expenditure than x. Thus, the probability to have an income higher than zero is 100% (the assumption is that everyone has some kind of income), for the first decile the probability that people have a higher income is 90%, and so forth. Let $x_1$, $x_2$,....$x_{10}$ be such that $x_1$ is the mean income for the first decile, $x_2$ is the mean income for the second decile, and $x_{10}$ is the mean income for the tenth decile. The set plotting the probability distribution would be G={(0,100%), ($x_1$, 90%), ($x_2$, 80%), ($x_3$, 70%), ($x_4$, 60%), ($x_5$, 50%), ($x_6$, 40%), ($x_7$, 30%), ($x_8$, 20%), ($x_9$, 10%), ($x_{10}$, 0%)}. We deemed a probability distribution to be robust if the annual value obtained for coefficient of determination is higher than 90%.

While most of the graphic representations are made using logarithmic values (log-log scale), we use normal values. However, logarithmic values of the same set yield similar values regarding the goodness of the fit measured using coefficient of determination.

The data considered in our paper are about Brazil [9], Philippine [10], Singapore [11], Sweden [12], Uganda [13], and the UK [14]. We want to highlight that most of the data here are about individual income, except for the UK where the data is about households. In the case of Uganda, we will explore how consumption, divided in deciles in a similar way as income, can be modelled using polynomial distribution.

## 4. Results

The analysis of the data using normal and not logarithmic values for graphical representation found that the best goodness of fit was in the case of polynomial distribution. This distribution seems optimal in the above mentioned context given the number of parameters, the high percentage for annual values obtained for the coefficient of determination. However, it isnoteworthy that in case of graphic representation using logarithmic values (log-log scale) we obtained similar values for the coefficient of determination. The most general formula we used was

$$y= P_1*x^2+ P_2*x + P_3 \quad (2)$$

which could be used as a first or a second degree polynomial depending on special particularity of each case. We present the values for the coefficients of the polynomials from fitting the data using Matlab Toolbox in the Appendices 1-6. The results were yielded for confidence intervals of 95 %. We exhibit graphically two relevant examples in the Figures 1 and 2. We chose specifically the following graphics as they have among the lowest values for coefficient of determination for all countries regarding income (Fig. 1) and expenditure (Fig. 2).

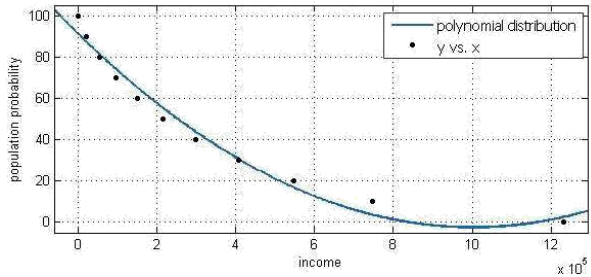

*Figure 1. The distribution of mean income for Philippine in the year 1997*

The equation describing the distribution is $Y=9.289*10^{-11}*x^2 + (-0.0001862)*x+90.89$, $R^2$=97.92 %

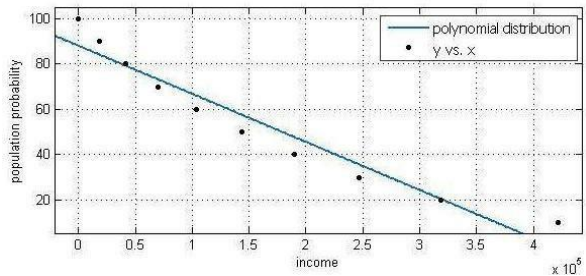

*Figure 2. The distribution of mean expenditure in the Uganda in the year 2006*

The equation describing the distribution is

$$Y = (-0.0002116)*x + 87.89, R^2=93.9 \%$$

The most important feature is the high robustness of the polynomial distribution in describing mean income and mean expenditure. The lowest value for coefficient of determination is 91.14% for income and 93.9% in case of expenditure. Further to our previous findings, we can see the applicability of the polynomial distribution to other macroeconomic systems. Its applicability is supported by several facts highlighting the diversity of these economies. First, we could apply it not only to developed countries with high income, but also to developing and underdeveloped countries characterized sometimes by low income. Second, the countries we took into account have very different economies from the point of view of their characteristics such as their model of development, evolution of macroeconomic indicators, and particularities of different national currencies. Also, brings additional evidence that polynomial distribution can successfully describe the income evolution for the entire income range, not only for the upper income segment of population which is believed to be described only by Pareto distribution.

The absolute novelty is that distributions from dynamic systems can be applied to an additional macroeconomic variable such as expenditure (or consumption). Moreover, polynomial distribution is robust in describing this variable. Brazil's case proves that different currencies do not change significantly the results for the same national economy at different points in time. Thus, the two currencies considered (new cruzeiro and reais) did not have a significant impact on the quality of analysis, the values for coefficient of determination remaining roughly the same throughout the years considered. In our analysis, we chose those polynomials which simultaneously can describe the probability distribution with values for coefficient of determination higher than 90% and which can have as little as possible coefficients for polynomials. This is the reason behind using the same variable for polynomials with different degrees in the cases of the countries considered. Based on the available data, we can conclude that polynomial distributions can have different degrees/forms according to the geographic space they describe based on some regional characteristics. Thus, countries from Asia and Latin America seem to be described by a second degree polynomial, while European and African countries seem to function with a first degree polynomial. Moreover, this occurs regardless the level of development of the countries analyzed. Time intervals of the data which span over many years point to a correlation between the values for the coefficient of determination and Gini coefficient. Thus, the values for coefficient of determination decrease in the time interval 1977-1991, while in the time interval 1992-2012 they increase and decrease without having a multiannual trend. Gini coefficient increases for the time interval 1977-1991 and afterwards increases and decreases without any overall trend [15]. These evolutions point to a negative relation between these indicators/indexes.

## 5. Conclusions

Polynomial distribution proved again its robustness in describing the income by applying it with high degree of success to a pool of countries having very different characteristics and which reacted very differently to the most recent world economic crisis. More importantly, we showed that different economic characteristics of the countries considered have an impact on the form of the polynomials used to fit the data.

Also, polynomial distribution proved its robustness in the analysis of expenditure/consumption.

The applicability of polynomial distribution in the analysis of expenditure/consumption opens the way to the analysis of this important macroeconomic variable. Also, it opens the way for other distributions from Physics in analysis of consumption. Moreover, the possibility to use the same statistical distribution in analysis of income and consumption may lead to a more rigorous theoretical analysis of the existing relation between income and consumption.

Also, this paper may shed light on the particularity of the income distribution mechanisms from countries having different social and economic characteristics.

# Appendix

*Appendix 1. Coefficients of the polynomials fitting mean income in Brazil*

| Year | P1 | P2 | P3 | $R^2$ (%) |
|---|---|---|---|---|
| 1960 | $3.809 \times 10^{-8}$ | -0.003728 | 89.53 | 97.73 |
| 1970 | $2.987 \times 10^{-8}$ | -0.00338 | 89.16 | 97.27 |
| 1980 | $2.987 \times 10^{-8}$ | -0.00338 | 89.16 | 97.27 |
| 1981 | $1.77 \times 10^{-5}$ | -0.08176 | 87.13 | 95.93 |
| 1992 | $1.869 \times 10^{-5}$ | -0.08339 | 86.44 | 95.75 |
| 2002 | $1.13 \times 10^{-5}$ | -0.0653 | 86.73 | 95.79 |

*Appendix 2. Coefficients of the polynomials fitting mean income in Philippine*

| Year | P1 | P2 | P3 | $R^2$ (%) |
|---|---|---|---|---|
| 1991 | $3.17 \times 10^{-10}$ | -0.0003439 | 91.81 | 98.4 |
| 1994 | $1.854 \times 10^{-10}$ | -0.0002618 | 91.94 | 98.47 |
| 1997 | $9.289 \times 10^{-11}$ | -0.0001862 | 90.89 | 97.92 |
| 2000 | $6.54 \times 10^{-11}$ | -0.0001559 | 90.97 | 97.99 |
| 2003 | $7.793 \times 10^{-11}$ | -0.0001698 | 91.36 | 98.19 |

*Appendix 3. Coefficients of the polynomials fitting mean income in Singapore*

| Year | P1 | P2 | P3 | $R^2$ (%) |
|---|---|---|---|---|
| 1980 | $8.327 \times 10^{-7}$ | -0.0173 | 89.53 | 97.79 |
| 1990 | $1.201 \times 10^{-7}$ | -0.006559 | 90.71 | 98.29 |
| 2000 | $4.871 \times 10^{-8}$ | -0.004124 | 87.61 | 97.16 |
| 2005 | $3.788 \times 10^{-7}$ | -0.01175 | 90.79 | 98.15 |
| 2006 | $3.366 \times 10^{-7}$ | -0.0111 | 90.96 | 98.2 |
| 2007 | $2.885 \times 10^{-7}$ | -0.01029 | 90.59 | 98.03 |

*Appendix 4. Coefficients of the polynomials fitting mean income in Sweden*

| Year | P1 | P2 | $R^2$ (%) |
|---|---|---|---|
| 2011 | -0.04188 | 87.01 | 94.61 |
| 2012 | -0.04126 | 87.18 | 94.75 |
| 2013 | -0.04047 | 87.2 | 94.73 |

*Appendix 5. Coefficients of the polynomials fitting mean expenditure in Uganda*

| Year | P1 | P2 | $R^2$ (%) |
|---|---|---|---|
| 2003 | -0.0002457 | 88.25 | 94.18 |
| 2006 | -0.0002116 | 87.89 | **93.9** |
| 2010 | -0.0001929 | 88.06 | 93.95 |

*Appendix 6. Coefficients of the polynomials fitting mean income in the UK*

| Year | P1 | P2 | $R^2$ (%) |
|---|---|---|---|
| 1977 | -0.0021 | 85.79 | 94.45 |
| 1978 | -0.002346 | 86.66 | 95.11 |
| 1979 | -0.002027 | 86.07 | 94.75 |
| 1980 | -0.001684 | 85.76 | 94.48 |
| 1981 | -0.001512 | 86.16 | 94.39 |
| 1982 | -0.00144 | 86.11 | 94.33 |
| 1983 | -0.001368 | 86.46 | 94.32 |
| 1984 | -0.001283 | 86.12 | 94.2 |
| 1985 | -0.001163 | 85.61 | 93.58 |
| 1986 | -0.001075 | 84.83 | 92.88 |
| 1987 | -0.0009737 | 84.31 | 92.5 |
| 1988 | -0.0008667 | 83.25 | 91.57 |
| 1989 | -0.0008072 | 83.33 | 91.87 |
| 1990 | -0.0007164 | 82.76 | 91.12 |
| 1991 | -0.0006672 | 83.03 | 91.4 |
| 1992 | -0.0006558 | 83.73 | 92.03 |
| 1993 | -0.0006441 | 83.75 | 91.74 |
| 1995 | -0.0006238 | 84.05 | 92.14 |
| 1996 | -0.0006094 | 84.46 | 92.49 |
| 1997 | -0.0005655 | 83.84 | 91.88 |
| 1998 | -0.0005338 | 83.63 | 91.72 |
| 1999 | -0.0005132 | 83.6 | 91.56 |
| 2000 | -0.0004868 | 83.1 | **91.14** |
| 2001 | -0.0004646 | 83.68 | 91.76 |
| 2002 | -0.0004302 | 83.21 | 91.18 |
| 2003 | -0.0004192 | 84.15 | 92.21 |
| 2004 | -0.0004109 | 84.2 | 92.29 |
| 2005 | -0.0003906 | 84.49 | 92.48 |
| 2006 | -0.00038 | 84.12 | 92.03 |
| 2007 | -0.0003623 | 84.12 | 0.92 |
| 2008 | -0.0003556 | 84.21 | 92.34 |
| 2009 | -0.0003478 | 84.16 | 92.18 |
| 2010 | -0.0003401 | 84.63 | 92.44 |
| 2011 | -0.0003561 | 85.21 | 92.65 |
| 2012 | -0.0003316 | 85.21 | 92.96 |